\documentclass[prx,twocolumn,showpacs,superscriptaddress]{revtex4}
\usepackage{amsmath}
\usepackage{amssymb}
\usepackage{amsthm}
\usepackage{amsfonts}
\usepackage{listings}
\usepackage{enumerate}
\usepackage{latexsym}
\usepackage{color}
\usepackage{bm}
\usepackage{hyperref}
\hypersetup{
 pdfnewwindow=true, colorlinks=true,
 linkcolor=blue, anchorcolor=blue,
 citecolor=blue, filecolor=blue,
 menucolor=blue, urlcolor=blue}

\newcommand{\beginsupplement}{%
        \setcounter{table}{0}
        \renewcommand{\thetable}{S\arabic{table}}%
        \setcounter{figure}{0}
        \renewcommand{\thefigure}{S\arabic{figure}}%
     }

\usepackage{psfrag}

\usepackage{bm}
\usepackage{graphicx}
\usepackage{subfigure}

\RequirePackage[normalem]{ulem} 
\RequirePackage{color}\definecolor{RED}{rgb}{1,0,0}\definecolor{BLUE}{rgb}{0,0,1} 

\newcommand{\beq}{\begin{equation}}
\newcommand{\eneq}{\end{equation}}

\newcommand{\ins}[1]{\;\;\text{#1}\;\;}

\def\bk{{\bf k}}
\def\bb{{\bf b}}
\def\br{{\bf r}}
\def\bn{{\bf n}}
\def\bp{{\bf p}}
\def\bg{{\bf g}}
\def\bt{{\bf t}}

\def\lms{Li$_{12}$Mg$_3$Si$_4$}
\newcommand{\wt}[1]{\widetilde #1}

\def\ie{{\it i.e.},\ }
\def\eg{{\it e.g.},\ }

\begin{document}

\tolerance 10000

\newcommand{\vk}{{\bf k}}

\draft

\title{Six-fold Excitations in Electrides}

\author{Simin Nie}
\affiliation{Beijing National Laboratory for Condensed Matter Physics,
and Institute of Physics, Chinese Academy of Sciences, Beijing 100190, China}
\affiliation{Department of Materials Science and Engineering, Stanford University, Stanford, California 94305, USA}

\author{B. Andrei Bernevig}
\email{bernevig@princeton.edu}
\affiliation{Department of Physics,Princeton University,Princeton, NJ 08544, USA}

\author{Zhijun Wang}
\email{wzj@iphy.ac.cn}
\affiliation{Beijing National Laboratory for Condensed Matter Physics,
and Institute of Physics, Chinese Academy of Sciences, Beijing 100190, China}
\affiliation{University of Chinese Academy of Sciences, Beijing 100049, China}

\date{\today}

\begin{abstract}
Due to the lack of full rotational symmetry in condensed matter physics, solids exhibit new excitations beyond Dirac and Weyl fermions, of which the six-fold excitations have attracted considerable interest owing to the presence of the maximum degeneracy in bosonic systems.
Here, we propose that a single linear dispersive six-fold excitation can be found in the electride \lms~and its derivatives. The six-fold excitation is formed by the floating bands of elementary band representation --- $A@12a$ --- originating from the excess electrons centered at the vacancies (\ie the $12a$ Wyckoff sites). 
There exists a unique topological bulk-surface-edge correspondence for the spinless six-fold excitation, resulting in trivial surface `Fermi arcs' but nontrivial hinge arcs. All energetically-gapped $k_z$-slices belong to a two-dimensional (2D) higher-order topological insulating phase, which is protected by a combined symmetry ${\cal T}{\wt S_{4z}}$ and characterized by a quantized fractional corner charge $Q_{corner}=\frac{3|e|}{4}$. Consequently, the hinge arcs are obtained in the hinge spectra of the $\wt S_{4z}$-symmetric rod structure. The state with a single six-fold excitation, stabilized by both nonsymmorphic crystalline symmetries and time-reversal symmetry, 
is located at the phase boundary and can be driven into various topologically distinct phases by explicit breaking of symmetries, making these electrides promising platforms for the systematic studies of different topological phases.
\end{abstract}

\maketitle

\section{introduction}
In the high-energy physics, Poincar\'{e} symmetry puts strong constraints on the standard model, leading to 
only three different types of fermions~\cite{pal2011dirac}, \ie Dirac, Weyl and Majorana fermions. Although several kinds 
of particles (such as protons) are confirmed to be Dirac fermions, the signature of Weyl and Majorana fermions is still lacking in particle physics experiments. In contrast to the stagnant situation in high-energy 
physics, great progresses have been made recently in the realization of their low-energy quasiparticles in condensed
matters~\cite{WangNa3Bi2012,WangCd3As22013,young2012dirac,tang2016dirac,Hua208prb,wan2011,xu2011Hgcr2se4,weng2015taas,huang2015weyl,wang2016prl,nie2017topological,wang2018large,nie2020prl,soluyanov2015type,nadj2014observation,he2017chiral,zhang2018observation,wang2018evidence,hourglass2017}.
\begin{figure}[!t]
\includegraphics[width=3.in]{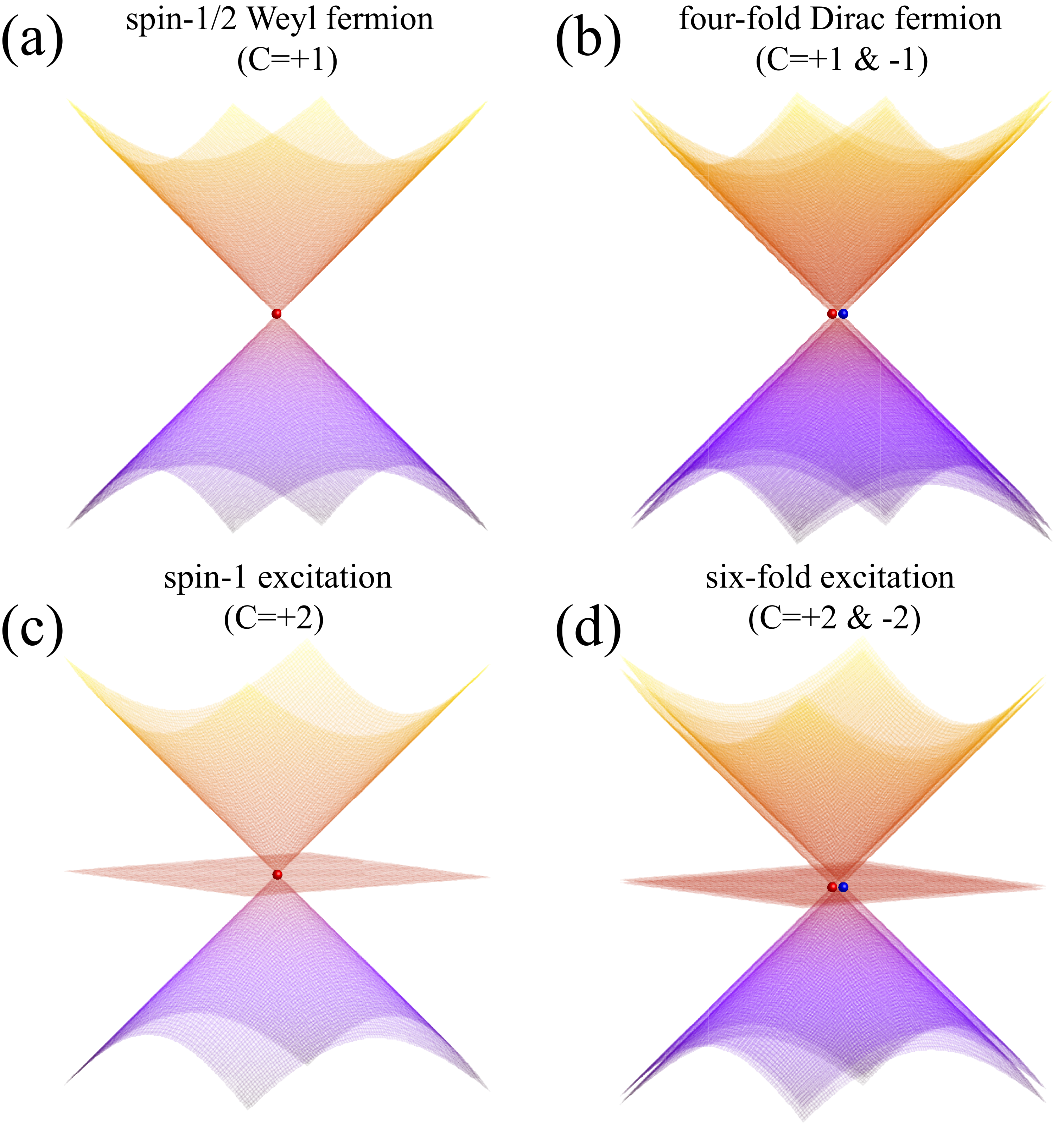}
     \caption{(color online). Schematics of different types of fermions. (a) Spin-1/2 Weyl fermion with charge $C=+1$. (b) Four-fold Dirac fermion. (c) Spin-1 excitation with charge $C=+2$. (d) Six-fold excitation.
}
\label{fig:fig1}
\end{figure}
More interestingly, due to the less constraints placed by the space group symmetries, condensed matter systems can host  
various new types of quasiparticles without counterparts in high-energy physics~\cite{bradlynaaf5037}, such as three-fold spin-1 excitations~\cite{bradlynaaf5037,CoSi,chang2017large,zhang2011prldoubleweyl}, four-fold spin-3/2 Rarita-Schwinger-Weyl (RSW) fermions~\cite{RSW}, and six-fold excitations~\cite{bradlynaaf5037}, {\it etc}. 
Similar to the four-fold Dirac fermions made up of two Weyl fermions with opposite chirality [Figs. \ref{fig:fig1}(a) and \ref{fig:fig1}(b)], the six-fold excitation is composed of two 
three-fold spin-1 excitations with opposite chirality [Figs. \ref{fig:fig1}(c) and \ref{fig:fig1}(d)].
Although the spin-1 excitations and spin-3/2 RSW fermions have been proposed in CoSi~\cite{CoSi} and verified in experiments~\cite{rao2019observation,CoSi_prl}, the six-fold excitations in the spinless systems have not been well investigated both in
experimental and theoretical consideration
due to the lack of ideal material candidates. Recently, the signature of spinful six-fold fermions [considering spin-orbit coupling (SOC)] has been observed in PdSb$_2$~\cite{adma201906046,sun2020,PRB201105,PhysRevB.99.161110}, but the presence of many trivial Fermi surfaces renders the problem very complicated for a simple understanding.
Unlike the spin-3/2 (or spin-1/2) fermions, the spin-1 excitations, and the two-fold Dirac fermions in 2D graphene, which cannot exist alone in the Brillouin zone (BZ), we show that a single linear dispersive six-fold excitation can be protected near the Fermi level in three-dimensional (3D) crystals.

\begin{figure*}[!t]
\includegraphics[width=6.2in]{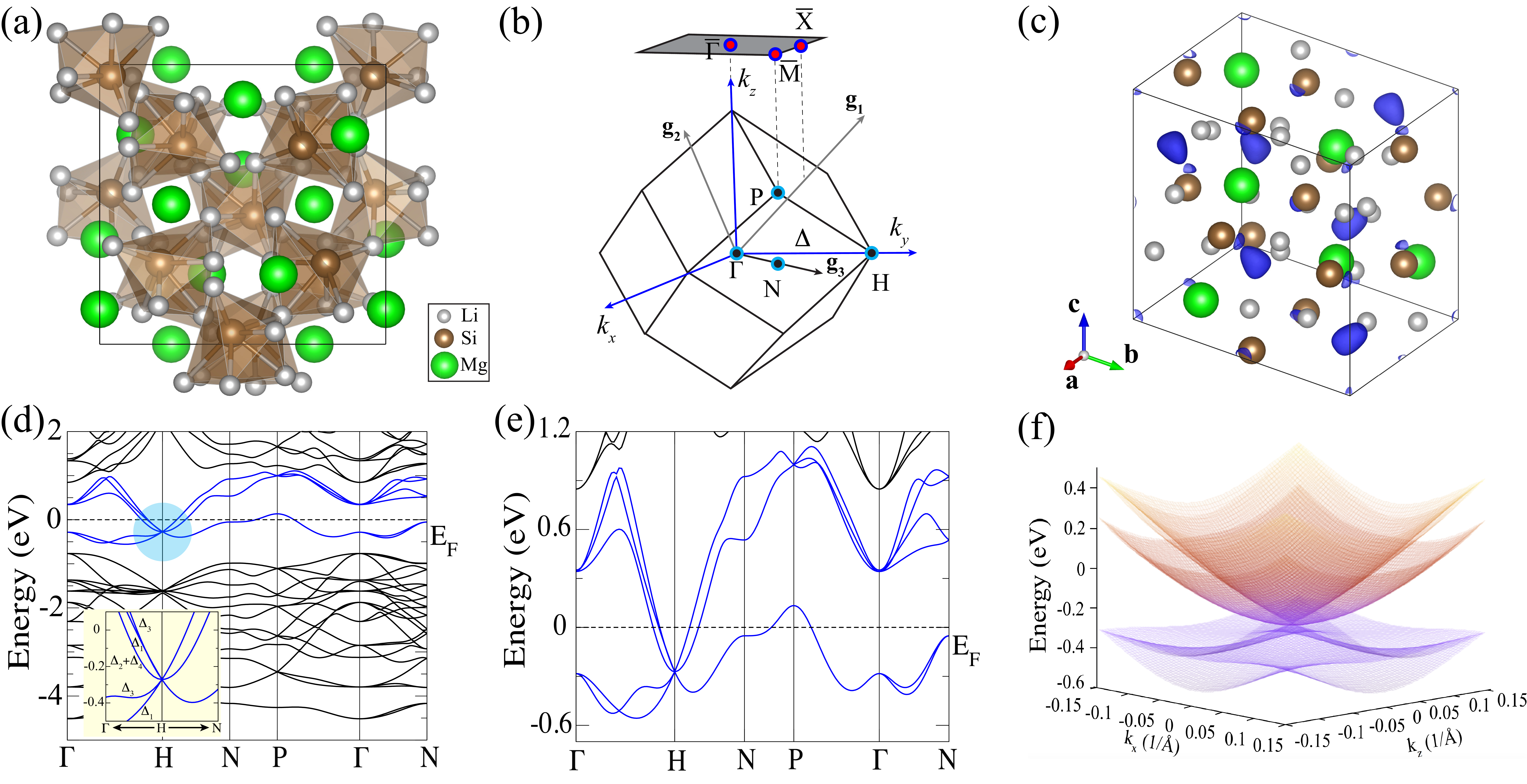}
      \caption{(color online). Crystal and electronic structures of \lms. 
     (a) Top view of the crystal structure of \lms. (b) Bulk BZ and (001) surface BZ with high symmetry points indicated. 
     (c) PED with an isosurface value of 0.0068 Bohr$^{-3}$ for the six blue-colored bands in Panel (d) or (e). The colored balls indicate the positions of atoms, while the irregular surfaces in blue indicate the charge distribution around the $12a$ sites.
    (d) and (e) The band structures of \lms~without SOC in different energy ranges. The inset in (d) shows the zoom-in band structure around the six-fold excitation.
     (f) 2D band structure in the $xoz$ plane around the six-fold excitation of the H point.
}
\label{fig2}
\end{figure*}

Recently, electrides have been found in inorganic crystals~\cite{dawes1986first,dye1987physical,matsuishi2003high,lee2013dicalcium}. They are defined as ionic crystals with excess electrons serving as anions confined in particular cavities. Compared with the electrons bounded by nuclei in atomic insulators (they in literature are classified as
trivial insulators with valence bands coming from \emph{real} atomic orbitals), there is a mismatch between the centers of excess valence electron density and atomic positions in the electrides~\cite{singh1993theoretical}. Due to the unbounded electronic states (called the floating bands), electrides have two unique properties: (i) electrides possess very low work functions, resulting in widely technological applications, including excellent electron emitters~\cite{phillips2000thermionic,huang1990low} and superior catalysts~\cite{kitano2012ammonia}; 
(ii) the floating bands usually are very close to the Fermi level (E$_\text{F}$), which are very likely to interact with other energy bands, resulting in a band inversion and nontrivial band topology~\cite{bradlyn2017topological,po2017symmetry}.
Because of the potential candidates of various topological states, electrides have received growing attention recently~\cite{murakami2018}.
Electrides Y$_2$C~\cite{huang2018topological,Y2C2018}, alkaline metal oxides~\cite{zhu2019computational} and Sr$_2$Bi~\cite{zhang2019topological} are predicted to be nodal-line semimetals. HfBr is a 3D topological insulator~\cite{Y2C2018}, while ferromagnetic LaCl and GdCl are nodal-line semimetal and quantum anomalous Hall insulator~\cite{LaClnie}, respectively.

In this letter, based on first-principles calculations and the low-energy effective $\bk\cdot \bp$ model, we demonstrate that 
the electride \lms~exhibits a single linear dispersive six-fold excitation at the H point 
near E$_\text{F}$, which is formed by the unbounded states of the excess electrons centered at the interstitial vacancies (\ie the $12a$ Wyckoff sites). These floating bands correspond to 
the elementary band representation (BR) of $A@12a$, in the theory of topological quantum chemistry (TQC)~\cite{bradlyn2017topological}. 
We also reveal the unique topological bulk-surface-edge correspondence for the spinless six-fold excitation, leading to trivial surface `Fermi-arc' states and topologically nontrivial hinge arc states. All energetically-gapped $k_z$-slices belong to a 2D higher-order topological insulating (HOTI) phase.
As a critical point of the phase transitions, this state can be driven into various topological phases by breaking specific symmetries.

\begin{table}[b]
\caption{
Wyckoff sites (WKS), site symmetries (Symm.), electron configurations (Conf.), irreps and the atom-orbital-induced BR (aBR) are listed for the crystal of \lms~in SG 220.
One can find that the total number of valence electrons is 52 (26 per spin).
}\label{table:site}
\begin{tabular}{c|c|c|c|cc|c}
\hline
\hline
Atom & WKS($\delta$) &  Symm.& Conf. &\multicolumn{2}{c|}{Irreps ($\rho$)}& aBR ($\rho@\delta$)\\
\hline
Li&$48e$& $1$   & $s^1$ & $s$&$:A$   &$A@48e$\\
     \hline
Mg&$12b$& $-4$  & $s^2$ & $s$&$:A$    & $A@12b$\\
     \hline
Si&$16c$& $3$   & $p^2$ & $p_z$&$:A_1$   & $A_{1}@16c$\\
\cline{5-7}   
     &      &  &       & $p_x,p_y$&$:^1$E$^2$E   & $^1$E$^2$E$@16c$\\
\hline
\hline
\end{tabular}
\end{table}

\section{results}

{\it Crystal structure and electronic structure.}---
Bulk \lms~crystallizes in a body-centered cubic structure with lattice constant $a=10.688$ $\text{\AA}$ and space group $I$-43d (SG 220)~\cite{kevorkov2004phase}, as shown in Fig.~\ref{fig2}(a).
The Li, Mg and Si atoms occupy $48e$, $12b$ and $16c$ Wyckoff sites, respectively, which are tabulated in Table~\ref{table:site}. Note that the multiplicities are given in a conventional cell, while the calculations are usually performed in a primitive cell.
The site symmetries, electron configurations and atomic-orbital-induced band representations (aBR) are presented in Table~\ref{table:site} as well.
Fig. \ref{fig2}(d) shows the band structure of Li$_{12}$Mg$_3$Si$_4$ without SOC along the high-symmetry $k$-points labeled in the first BZ of Fig.~\ref{fig2}(b) [see calculation details in Section \ref{sup:A} of the Supplementary Material (SM)]. 
One can notice that a set of 24 bands in the energy range $-6$ eV to $-0.6$ eV are well separated from the other bands. Besides, there are two valence bands right below E$_\text{F}$, which are connecting to four conduction bands at a single point H. 
These six bands near E$_\text{F}$ are colored in blue in Fig. \ref{fig2}(d), which are actually isolated from high-energy bands, as shown in Fig. \ref{fig2}(e). In the following, we will show that the six blue-colored bands do not belong to an aBR.

The analysis of (elementary) BRs in TQC is an effective method to obtain both band structure topology in momentum space and orbital characters in real space.
A BR of $\rho@\delta$ is formed by the states of the $\rho$ irrep centered at the $\delta$ Wyckoff site. 
In other words, it can diagnose not only the nontrivial topology of electronic band structures, but also the centers of electron density for a set of bands. 
Thus, the mismatch between the electron density centers and the atomic positions in crystals can be well diagnosed by the TQC theory~\cite{todo2020}.  
The irreps of the 30 electronic states at high-symmetry $k$-points are obtained by the $irvsp$ program~\cite{gao2020irvsp} and are shown in Table~\ref{table:irrep}. Comparing them with the BRs in TQC listed on the Bilbao Crystalline Server (BCS)~\cite{BCSserver}, we find that as  expected, the 24 lower bands belong to the BRs of $(A_1\oplus ^1$E$^2$E)$@16c$, originating from $p$ orbitals of Si atoms. Interestingly, we also find that the six blue-colored bands belong to the BR of $A@12a$, which is not an aBR because of the un-occupancy of the $12a$ Wyckoff sites. Thus, one can conjecture that the ionic compound \lms~can be an electride with excess electrons serving as anions at the vacancies of the $12a$ sites in terms of the representation theory.
Furthermore, we calculate the partial electron density (PED) for these six bands. The results shown in Fig. \ref{fig2}(c) indicate the existence of a charge distribution around the $12a$ sites~\cite{PED}, agreeing well with the analysis of BRs.

\begin{table}[!t]
\caption{
Irreps and BRs for 30 bands in the energy range $-6.0$ eV to 1.0 eV in Li$_{12}$Mg$_3$Si$_4$.
Their irreps are given in order of increasing energy eigenvalues.
}\label{table:irrep}
\begin{tabular}{p{1.2cm}|p{1.25cm}|p{1.35cm}|p{0.93cm}|p{1cm}|p{1.8cm}}
\hline
\hline
Bands & $\Gamma$ (GM) & H & P &N & BRs\\
\hline
lower 24 bands &  GM3 (2) GM5 (3) GM2 (1) GM5 (3) GM3 (2) GM4 (3) GM5 (3) GM4 (3)     GM1 (1) GM4 (3)
&  H4H5  (6) H4H5  (6) H3H3  (4) H4H5  (6) H1H2  (2)
&  P2  (2) P3  (4) P1  (2) P3  (4) P3  (4) P1  (2) P2  (2) P3  (4) 
&  N1  (2) N1  (2) N1  (2) N1  (2) N1  (2) N1  (2) N1  (2) N1  (2)  N1  (2) N1  (2) N1  (2) N1  ~(2) &  $(A_1+^1$E$^2$E) ~~~$@16c$\\
\hline
6 bands near $E_F$ &  GM3 (2) GM1 (1) GM5 (3)
&  H4H5 (6)
&  P1  (2) P3 (4)
&  N1  (2) N1  (2) N1~ (2) &{\color{blue} $A@12a$}\\
\hline
\hline
\end{tabular}
\end{table}

Given that the conducting electrons are crucial to the physical properties of materials, we focus on the six bands of the $A@12a$ BR near E$_F$, to obtain the nature of the conducting electrons, especially the six-fold degenerate state at H. 
The linear dispersive bands around the H point resemble a six-fold excitation [highlighted in a shadowed circle in Fig. \ref{fig2}(d)], whose band dispersions near the H point are shown in the inset of Fig. \ref{fig2}(d).
In Weyl semimetals, Weyl nodes are coming in pairs with opposite chirality. In 2D graphene, two-fold Dirac fermions are also formed at two non-equivalent K points. However, the state with a \emph{single} linear dispersive spinless six-fold excitation in \lms~is unique, which has not been proposed before.  This state is actually located at the phase boundary and protected by both nonsymmorphic crystalline symmetries and time-reversal symmetry (TRS).

{\it Symmetry analysis and six-fold excitations.}---
First, along the $\Gamma$--H line in Fig.~\ref{fig2}(b) [\ie the [100] ($\Delta$) direction], the six-fold degenerate bands split into four singly-degenerate bands and one doubly-degenerate band, whose irreps can be labeled by the point group $C_{2v}$, as shown in the inset of Fig.~\ref{fig2}(d).
Second, the six-fold degenerate bands split into three two-fold degenerate bands along the H--N line ([110] direction) with the glide mirror symmetry $\wt M_{110}[\equiv\{I C_{2,110}|\frac{1}{2},1,1\}]$, which is on the surface of the first BZ. 
Hereafter, $I$ is spatial reflection, $C_{m,\bn}$ is a rotation by 360$^{\circ}/m$ about the $\bn$ axis, and the translation ($x_1,x_2,x_3$) is given in units of the primitive lattice vectors ($\bt_1,\bt_2,\bt_3$) defined in Section \ref{sup:B} of the SM. In fact, all the bands along the H--N line are doubly degenerate due to the combined anti-unitary symmetry of $\wt M_{1-10} [\equiv\{I C_{2,1\bar 10}|\frac{1}{2},-\frac{1}{2},\frac{1}{2}\}]$ and TRS ($\cal T$), satisfying the relation $({\cal T} \wt {M}_{1-10})^2=\{E|0,0,1\}=-1$ on the H--N line.

Then, we investigate the symmetry protection of the six-fold excitation and derive the effective $\bk \cdot \bp$ Hamiltonian at H point. The nonsymmorphic little group of the H point has three unitary generators $\{I C^{-1}_{4x}|\frac{1}{2},0,0\}$, $\{I C_{2,110}|\frac{1}{2},0,0\}$ and $\{C^{-1}_{3,1\bar 1\bar 1}|1,\frac{1}{2},\frac{1}{2}\}$, and an anti-unitary operator TRS. Under its little group, the irreps of the six bands at H are H4H5 (H4 and H5 stick together due to TRS), which are assigned by the $irvsp$ package in the convention of the BCS notation~\cite{gao2020irvsp,stokes2013tabulation}. 
Under the basis of H4H5 irreps, the matrix representations of the four symmetries are given in Section \ref{sup:C} of the SM, explicitly.

\begin{figure}[!b]
\includegraphics[width=3.5in]{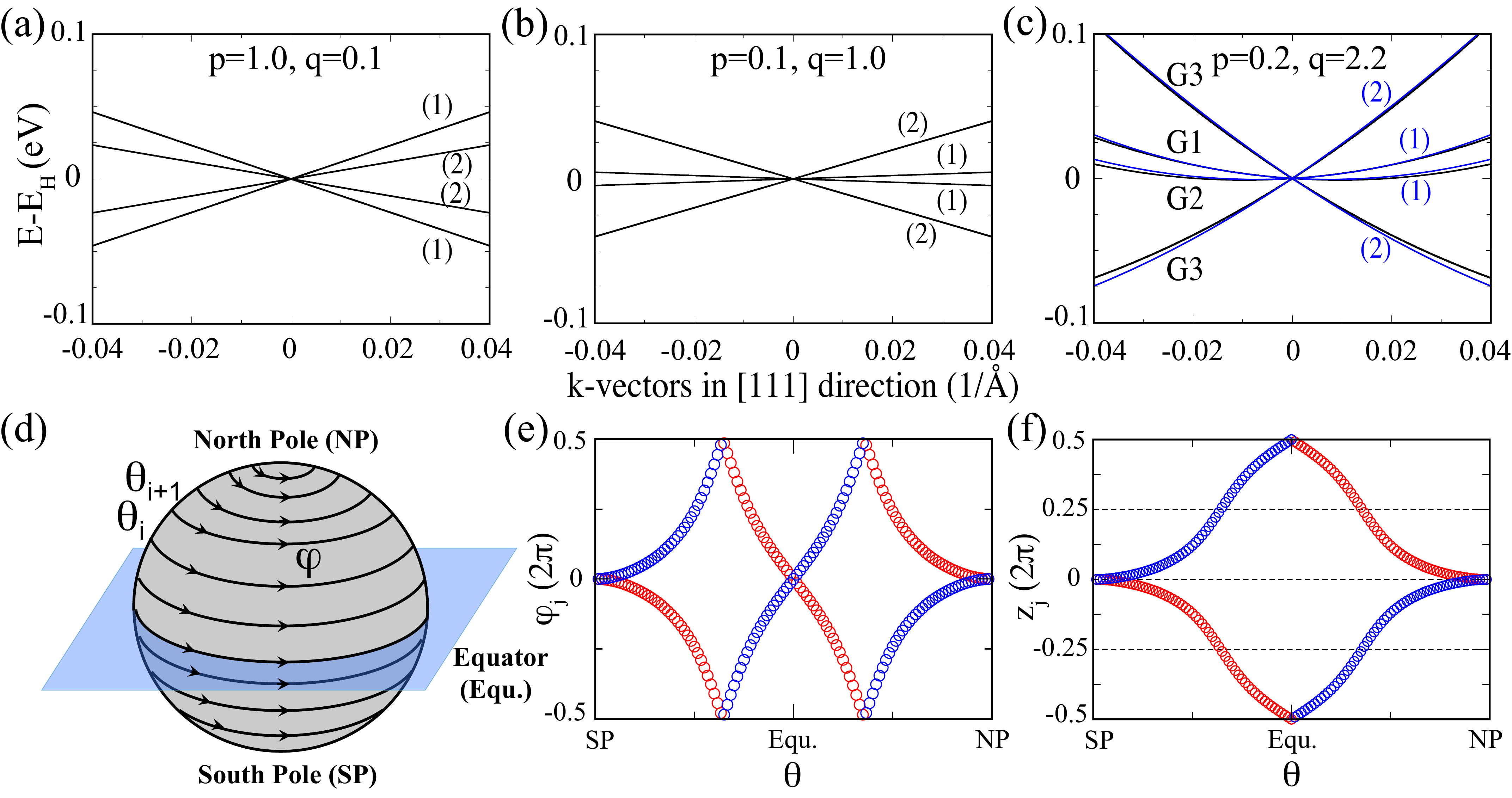}
     \caption{(color online). Band dispersions of the $\bk\cdot \bp$ model and Wilson loop spectra for two lowest occupied bands. The band dispersions of the low-energy effective model along [111] direction with $p=1.0$, $q=0.1$ (a); $p=0.1$, $q=1.0$ (b); and $p=0.2$, $q=2.2$ (c) in units of eV$\cdot$\AA. The black and blue lines are obtained from low-energy effective $\bk\cdot \bp$ model and first-principles calculations, respectively. The energy of the six-fold excitation at H is chosen as the reference energy (E$_\text{H}=0$ eV). 
(d) A series of 1D Wilson loops on a closed sphere enclosing the six-fold excitation. (e) The Wannier charge centers for $\theta_j$ loops schematically shown in (d).
(f) The phases of the $\overline C_{2z}$ eigenvalues ($e^{-iz_j}$) for the Wilson bands in (e).
}
\label{fig3}
\end{figure}

To construct the low-energy effective $\bk\cdot \bp$ Hamiltonian in the vicinity of the H point ($\bk_0$) in terms of $\delta \bk\equiv\bk -\bk_0$, our strategy is to find the symmetry-invariant matrix function ${\cal H}(\delta \bk)$ satisfying the condition placed by the symmetry $\cal O$:
\beq
{\cal O}{\cal H}(\delta \bk) {\cal O}^{-1}= {\cal H}(g_{\cal O} \delta \bk)
\eneq
After the consideration of all symmetry restrictions, the low-energy effective model 
can be up to the first order of $\delta \bf{k}$ written as 
\begin{widetext}
\begin{eqnarray}
{\cal H}(\delta \bk)&=
\left(
\begin{array}{cccccc}
 0 & p \delta k_x & -p \delta k_y & 0 & q^* \delta k_x & q^* \delta k_y \\
 p \delta k_x & 0 & -p \delta k_z & -q^* \delta k_x & 0 & -q^* \delta k_z \\
 -p \delta k_y & -p \delta k_z & 0 & -q^* \delta k_y & q^* \delta k_z & 0 \\
 0 & -q \delta k_x & -q \delta k_y & 0 & -p \delta k_x & p \delta k_y \\
 q \delta k_x & 0 & q \delta k_z & -p \delta k_x & 0 & p \delta k_z \\
 q \delta k_y & -q \delta k_z & 0 & p \delta k_y & p \delta k_z & 0  \\
\end{array}
\right) &=
\left(
\begin{array}{cccccc}
p H_{3}(0,\delta \bk') & -iq^*H_{3}(\pi/2,\delta \bk') \\
 i qH_{3}(\pi/2,\delta \bk')& -pH_{3}(0,\delta \bk')  \\
\end{array}
\right) \\
& \ins{ where }\delta \bk'\equiv (\delta k_x,-\delta k_y,-\delta k_z),
~~~~~~ H_{3}(\phi,\delta \bk) &=
\left(
\begin{array}{ccc}
 0 & e^{i \phi } \delta k_x & e^{-i \phi } \delta k_y \\
 e^{-i \phi } \delta k_x & 0 & e^{i \phi } \delta k_z \\
 e^{i \phi } \delta k_y & e^{-i \phi } \delta k_z & 0 \\
\end{array}
\right)
\end{eqnarray}
\end{widetext}
where $p$ has to be a real parameter and $q$ is an arbitrary parameter. When the value $\phi=\pi/2$, the Hamiltonian $H_3$ takes the form of $H_3(\pi/2,\delta \bk)=\delta \bk\cdot \bf S$, where the matrices $\bf S$ are the generators of the rotation group SO(3) in the spin-1 representation. 
In the limit of $|q|>>|p|$ (resp. $|p|>>|q|$), this six-fold degenerate state consists of two three-fold spin-1 excitations (resp. two three-fold excitations of $H_3(0,\delta \bk')$), which are related by TRS. 
In the two limits, the band splittings are also different in the [111] direction, as shown in Figs. \ref{fig3}(a) and  \ref{fig3}(b). By fitting the energy bands of \lms~with a more general form of $\alpha\delta \bk^2+{\cal H}(\delta \bk)$, the parameters $\alpha$, $p$ and $q$ are obtained to be 12 eV$\cdot$\AA$^2$,  $0.2$ eV$\cdot$\AA~and $2.2$ eV$\cdot$\AA, respectively. 
The dispersions along the [111] direction are presented in Fig.~\ref{fig3}(c), while those along [100] and [110] directions are given in Fig.~\ref{figs1} in Section \ref{sup:D} of the SM.
Therefore, the spinless six-fold degenerate state in \lms~is in the limit of two spin-1 excitations with opposite chirality (\ie the six-fold excitation), leading to the net topological charge of the two lowest bands being zero. 
It is worth noting that the state is different from the spinful six-fold excitations in SGs 205, 206 and 230 with double degeneracy in all directions, 
and those in SGs 198, 212 and 213 consisting of two spin-1 excitations with the same chirality~\cite{bradlynaaf5037}.

{\it Wilson loop technique.}---
Although the net topological charge of the two lowest bands is zero, we can still define a nontrivial Wilson loop spectrum on a spherical surface enclosing the six-fold excitation at the H point, \ie $(\frac{1}{2},\frac{1}{2},-\frac{1}{2})$ [given in units of the primitive reciprocal vectors ($\bg_1$, $\bg_2$ and $\bg_3$) defined in Section \ref{sup:B} of the SM], to confirm the topological nature of the six-fold excitation. Here we consider the family of Wilson loop matrices $W_\theta$, parameterized by the azimuthal angle $\theta$, as shown in Fig.~\ref{fig3}(d).
By plotting the phases $\varphi_j$ of the individual eigenvalues ($e^{i\varphi_j}$) of the matrices $W_\theta$ as a function of $\theta$ in Fig.~\ref{fig3}(e) (also called Wilson bands for short), 
one can find that even though the net topological charge is zero, the two Wilson bands do wind by $\pm 4\pi$.
Next, we will prove that the crossings of Wilson bands are protected and robust against adding more trivial occupied bands.

First, we are considering the robustness of the crossings of Wilson bands at $\varphi_j=\pi$.
In view of the rotation symmetry $\wt C_{2z}\equiv\{C_{2z}|\frac{1}{2},0,\frac{1}{2}\}$ satisfying the relation $[\wt C_{2z}]^2=\{E|000\}=1$, we have
\beq
\begin{split}
\wt C_{2z} W_{\theta,\frac{5\pi}{2}\leftarrow \frac{\pi}{2}}\wt C_{2z}^{-1}&=W_{\theta,\frac{7\pi}{2}\leftarrow \frac{3\pi}{2}}\\
&=W_{\theta,\frac{3\pi}{2}\leftarrow \frac{\pi}{2}}W_{\theta,\frac{5\pi}{2}\leftarrow \frac{\pi}{2} }W^\dagger_{\theta,\frac{3\pi}{2}\leftarrow \frac{\pi}{2}}\\
\end{split}
\eneq
where $W_{\theta,B\leftarrow A}$ represents a parallel transport from $\phi=A$ to $\phi=B$ on the $\theta$-loop of ($k_r^{\theta} \cdot cos[\phi],k^{\theta}_r \cdot sin[\phi], k^{\theta}_z$) with the $xy$-plane radius $k_r^{\theta}$.
Then we can redefine the symmetry as
$\overline{C}_{2z}\equiv W_{\theta,\frac{3\pi}{2}\leftarrow \frac{\pi}{2}}^{-1} \wt C_{2z}$, and find
\beq
\overline C_{2z} W_{\theta,\frac{5\pi}{2}\leftarrow \frac{\pi}{2}}\overline C_{2z}^{-1}=W_{\theta,\frac{5\pi}{2}\leftarrow \frac{\pi}{2}}
\eneq
which means that the Wilson bands can actually be labeled by the $\overline C_{2z}$ eigenvalues
\beq
\begin{split}
[\overline{C}_{2z}]^2&=W_{\theta,\frac{3\pi}{2}\leftarrow \frac{\pi}{2}}^{-1} \wt C_{2z}W_{\theta,\frac{3\pi}{2}\leftarrow \frac{\pi}{2}}^{-1} \wt C_{2z}\\
&=W_{\theta,\frac{3\pi}{2}\leftarrow \frac{\pi}{2}}^{-1} W_{\theta,\frac{5\pi}{2}\leftarrow \frac{3\pi}{2}}^{-1} \wt C_{2z} \wt C_{2z}\\
&=W_{\theta,\frac{5\pi}{2}\leftarrow \frac{\pi}{2}}^{-1} [\wt C_{2z}]^2\\
\overline{C}_{2z}&=\xi_j e^{-i\varphi_j/2}=e^{-iz_j},~ z_j\equiv i \text{log}(\xi_j)+\varphi_j/2\\
\end{split}
\eneq
where $\xi_j$ ($\pm 1$) are the $\wt C_{2z}$ eigenvalues of the occupied bands at the south pole (SP) and north pole (NP), which are on the $z$-axis. 
After we plot the phases $z_j$ as a function of $\theta$ in Fig.~\ref{fig3}(f), we find that the two Wilson bands at $\varphi_j=\pi$  actually have different $\overline{C}_{2z}$ eigenvalues (\ie $e^{-iz_j}=\pm i$). 

Then, we are considering the robustness of the crossing of Wilson bands on the equator ($equ.$), \ie $\theta=equ.$, by 
introducing a combined anti-unitary symmetry of TRS and $\wt M_{110}[\equiv\{IC_{2,110}|\frac{1}{2},1,1\}]$.
\beq
\begin{split}
\wt M_{110} W_{equ.,\frac{5\pi}{2}\leftarrow \frac{\pi}{2}}\wt M_{110}^{-1}&=W^{\dagger}_{equ.,\frac{7\pi}{2}\leftarrow \frac{3\pi}{2}}\\
{\cal T}\wt M_{110} W_{equ.,\frac{5\pi}{2}\leftarrow \frac{\pi}{2}}({\cal T}\wt M_{110})^{-1}&=W^\dagger_{equ.,\frac{5\pi}{2}\leftarrow \frac{\pi}{2}}.\\
\end{split}
\eneq
The Wilson Hamiltonian $H_W(\theta)$ is defined as 
\begin{eqnarray}
&W_{\theta,\frac{5\pi}{2}\leftarrow \frac{\pi}{2}} \equiv e^{iH_{W}(\theta)} \\
&{\cal T}\wt M_{110} H_W(equ.)({\cal T}\wt M_{110})^{-1}=H_W(equ.)
\end{eqnarray}
Since $[{\cal T}\wt M_{110}]^2=-1$, the eigenvalues ($\varphi_j$) of $H_W(equ.)$ are doubly degenerate.

Therefore, we conclude that the crossing points at $\varphi_j=\pi$ are protected by the $\overline{C}_{2z}$ symmetry, while the crossing point at $\theta=equ.$ is protected by the anti-unitary symmetry ${\cal T}\wt M_{110}$. The winding feature of $\pm 4\pi$ is protected and robust against adding more trivial occupied bands (see more details in Section \ref{sup:E} of the SM).

\begin{figure}[t]
\includegraphics[width=3.4in]{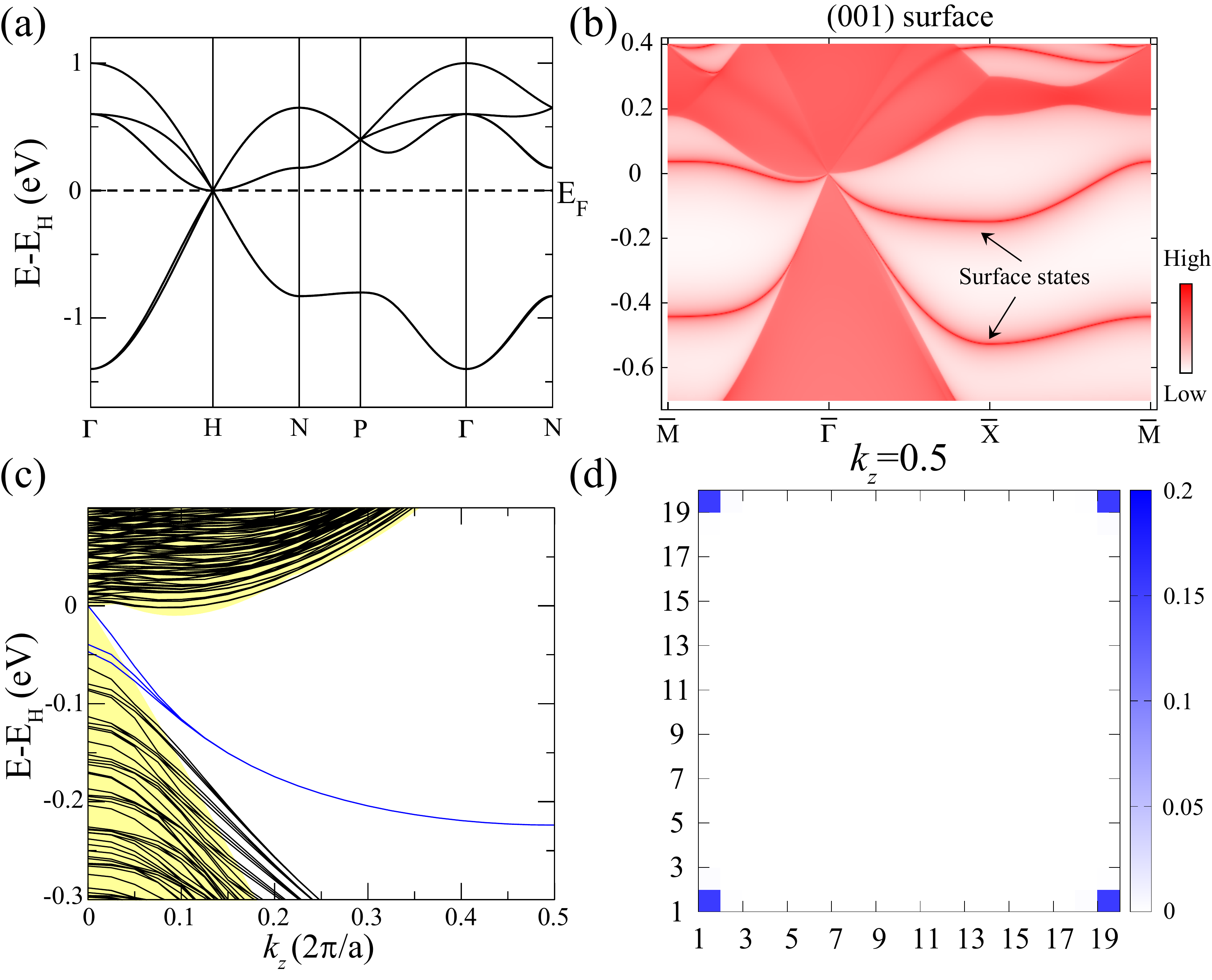}
\caption{(color online). 
The bulk and boundary states of the TB Hamiltonian. 
The bulk bands (a), (001) surface dispersions (b), and [001] hinge dispersions (c). The yellow shadow denotes the projections of the bulk states.  (d) The real-space distributions of the four hinge states for $k_z=\frac{\pi}{a}$ on the structure of 1D $20\times20$ supercell.
}
\label{fig:fig4}
\end{figure}

{\it Tight-binding model, surface states and hinge arc states.}---
The six-fold excitation can be easily reproduced by a tight-binding (TB) model with an $s$-orbital at each $12a$ Wyckoff site. It reads as follows
\begin{eqnarray}
&H_{TB}(\bk)=\epsilon_0+\sum_{|\br|=d_{j=1,2,3}} t_j e^{i\bk\cdot \br} 
\label{tb}
\end{eqnarray}
where $\epsilon_0$ is the on-site energy of the $s$-orbital, $t_{1,2,3}$ are the nearest-, next-nearest-, and next-next-nearest-neighbor hopping parameters with the distances $d_{1,2,3}$, respectively. The parameters of the TB band structure plotted in Fig.~\ref{fig:fig4}(a) are given in Table~\ref{table:tb} in Section \ref{sup:F} of the SM (SM \ref{sup:F}). A single six-fold excitation emerges at the H point, whose dispersions in all directions are qualitatively consistent with the $\bk\cdot \bp$ model and  
first-principles calculations.
By using the Green's function on a semi-infinite structure, the (001) surface dispersions are obtained in Fig.~\ref{fig:fig4}(b), where two surface states are terminated at the projection of the six-fold excitation (\ie $\overline{\Gamma}$).
Similar to the four-fold Dirac points, the six-fold excitations can yield trivial `Fermi arcs' on the surfaces (unless they are finely tuned). Accordingly, this kind of `Fermi-arc' surface states (\ie their starting point and ending point coincide) are usually found in the semimetals with Dirac fermions (\eg Na$_3$Bi) or six-fold excitations [\eg \lms,~as shown in Figs.~\ref{figs3}(a) and \ref{figs4}(a)].

Even though the surface states are not topologically protected in \lms, there are topologically nontrivial hinge arc states on its [001] hinges, as analogous to the hinge arcs in the Dirac semimetals. For the Dirac semimetals, the energetically-gapped 2D $k_z$-slices are classified into two different phases (one is a trivial phase, and the other one is a nontrivial HOTI phase), which are separated by the Dirac points [\ie $(0,0,\pm k_z^c)$]~\cite{Ben2020nc}. Interestingly, in the semimetals with a single spinless six-fold excitation, all energetically-gapped 2D $k_z$-slices belong to a HOTI phase, except for the gapless plane with the six-fold excitation.
The 2D nontrivial HOTI phase is closely related to a \emph{filling anomaly}~\cite{anomaly}: a mismatch between the number of electrons required to simultaneously satisfy charge neutrality and the crystal symmetry (\ie ${\cal T}\wt S_{4z}$). 
By computing one-dimensional (1D) Wilson loops along the $\bb_1^*$ ($\equiv\bg_3$) and $\bb_2^*$ ($\equiv \bg_1-\bg_2$) directions, our results show that the two charge centers are quantized to [0.25$\bb_1$,0.25$\bb_2$] and [0.75$\bb_1$,0.75$\bb_2$] with $\bb_1\equiv \{a/2, a/2\}$ and $\bb_2\equiv \{-a/2, a/2\}$, which correspond to $\{0, 0.25a\}$ and $\{0, 0.75a\}$, respectively, in Cartesian coordinates in the projected $xy$-plane. The quantizations are protected by the ${\cal T}\wt S_{4z}$ symmetry (with $\wt S_{4z}\equiv\{IC^{-1}_{4z}|00\frac{1}{2}\}$) for each energetically-gapped 2D $k_z$-slice. Because the two obtained Wannier charge centers are two distinct ${\cal T}\wt S_{4z}$-invariant positions, it is impossible to have any choice of Wannier center assignment that preserves charge neutrality and $\wt S_{4z}$ symmetry simultaneously~\cite{anomaly}.

Unlike the chiral/helical hinge modes connecting the conduction bands and valence bands in the 3D higher-order topological insulators, the in-gap hinge arc states are connecting the projection of the six-fold excitation in the hinge spectra of \lms. 
In Fig.~\ref{fig:fig4}(c), we plot the  [001] hinge dispersions of an 1D $\wt S_{4z}$-invariant 20$\times$20 supercell with open-boundary conditions in $\bb_1$ and $\bb_2$ directions. The in-gap bands colored in blue are the four hinge states, which are well localized on four hinges. 
The real-space distributions of the four hinge states for $k_z=\frac{\pi}{a}$ are shown in Fig.~\ref{fig:fig4}(d) (see more details in Section \ref{sup:G} of the SM).
Three electrons occupy the four hinge states in a nonzero $k_z$ plane ($k_z\neq 0$), resulting in a fractional corner charge $Q_{corner}=\frac{3|e|}{4}$. Therefore, the [001]-hinge arc states, well separated from the bulk and surface projections except for the projection ($k_z=0$) of the six-fold excitation, confirm the nontrivial bulk topology in \lms.

\begin{figure}[t]
\includegraphics[width=3.4in]{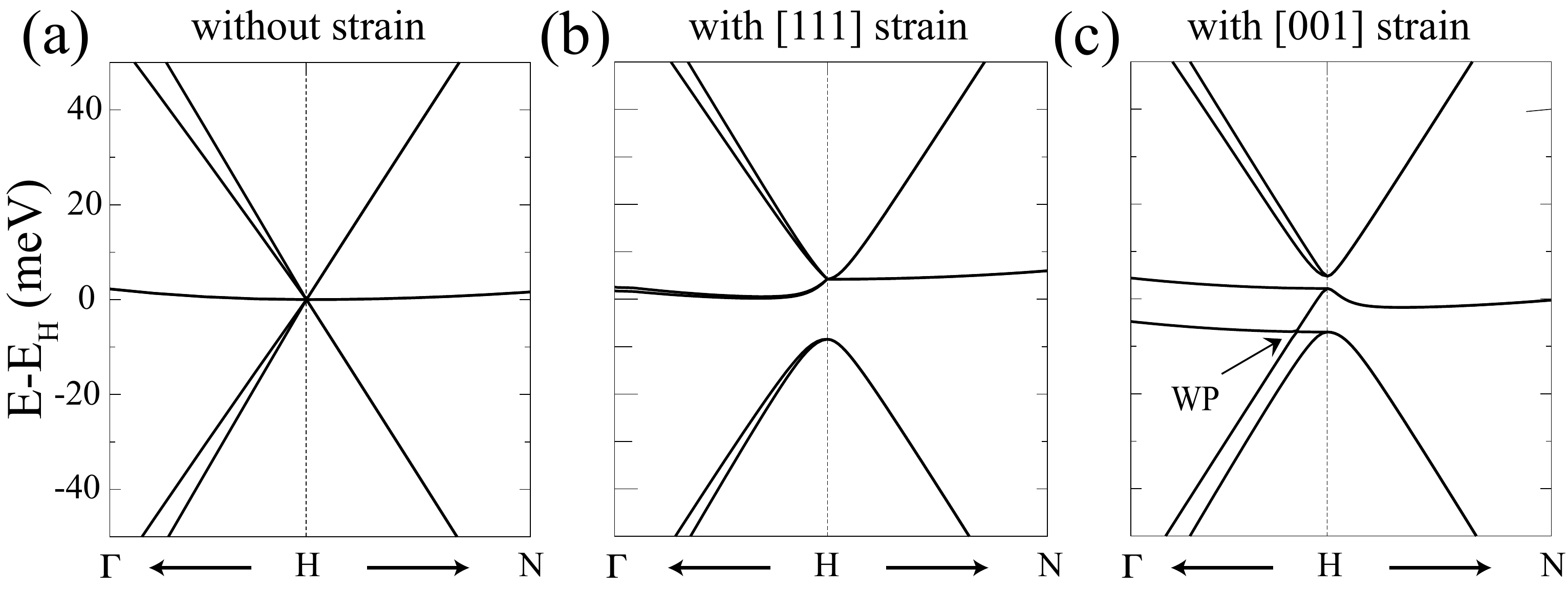}
\caption{(color online). 
The bulk band structures without strain (a), with [111] strain (b), and with [001] strain (c). 
}
\label{fig:fig5}
\end{figure}

{\it Phase transitions in \lms.}---
The state with a single six-fold excitation can be driven into various topological states after consideration of strain effect, which can be simulated through multiplying the hopping parameters in the Hamiltonian by a factor of $|\br'/\br|$ with $\br~(\br')$ the distances of the hoppings without (with) strains. Here the compressive strains along [001] and [111] directions are studied, and the band dispersions near H are shown in Figs.~\ref{fig:fig5}(a-c). 
When the strain is applied in [001] direction, the sixfold excitation splits into four Weyl points on the $k_z=0$ plane, whose projections on the (001) surface are the ends of two long visible `Fermi-arc' states extending through the BZ in spite of the small spacial splitting of Weyl points, as shown in Figs. \ref{figs3}(b,c) and \ref{figs4}(b,c) in SM \ref{sup:F}. 
On the other hand, once strain in [111] direction is applied, a phase transition from a single six-fold excitation to two sets of nodal lines is observed. Each set of nodal lines consists of one line segment in [111] direction and three $C_{3,111}$-related arc segments, as shown in Figs.~\ref{figs5}(b,c) in SM \ref{sup:F}. The nodal line in [111] direction is protected by a two-fold irrep with the three-fold rotation $C_{3,111}$, while the nodal lines on the (1$\bar{1}$0), ($\bar{1}$01) and (01$\bar{1}$) planes are protected by $\wt M_{1-10}$, $\wt M_{-101}$ and $\wt M_{01-1}$, respectively [Fig.~\ref{figs5}(c)].

\section{Discussion}
In summary, the electride \lms~hosting a single linear dispersive six-fold excitation near $E_F$ can be well understood by the analysis of BRs in TQC. 
The six-fold excitation is formed by the floating states of the unbounded electrons in \lms, labeled by the BR of $A@12a$ with an s-like symmetry (\ie $A$ irrep) at the $12a$ Wyckoff sites. Given the unique topological bulk-surface-edge correspondence for the excitation, 
surface `Fermi-arc' states and topologically nontrivial hinge arc states are obtained. 
When strain effect is considered, the six-fold excitation can be driven into Weyl points or nodal lines {\it etc}.  
Due to the negligible strength of SOC, a twelve-fold degenerate excitation (the highest degeneracy in electronic systems) can be found in \lms~after the consideration of the spin degrees of freedom (see Section \ref{sup:H} of the SM).
Moreover, there are a rich family of $A_{12}A'_{3}B_4$ compounds, including Li$_{12}$Mg$_3$Si$_4$,  Li$_{12}$Al$_3$Si$_4$ and Li$_{15}$Ge$_4$ {\it etc}. (see Section \ref{sup:H} of the SM), making the chemical potential of \lms~easily tunable. These electrides can serve as new platforms for the study of the topological semimetallic states and phase transitions in future experiments.

\begin{acknowledgments}
This work was supported by the National Natural Science Foundation of China (Grants No. 11974395, No. 11504117), the Strategic Priority Research Program of Chinese Academy of Sciences (Grant No. XDB33000000), and the CAS Pioneer Hundred Talents Program.
Z.W. and B.A.B. were supported by the Department of Energy Grant No. DE-SC0016239. B.A.B. was additionally supported by the National Science
Foundation EAGER Grant No. DMR 1643312, Simons Investigator Grant No.
404513, ONR Grant No. N00014-14-1-0330, the Packard Foundation, the
Schmidt Fund for Innovative Research, and a Guggenheim Fellowship from
the John Simon Guggenheim Memorial Foundation.
\end{acknowledgments}

\bibliography{lms}

\clearpage
\begin{widetext}

\beginsupplement{}
\setcounter{section}{0}
\section*{SUPPLEMENTARY MATERIAL}
\subsection{Calculation method}
\label{sup:A}

Our first-principles calculations are carried out by the projector augmented wave method implemented in the
Vienna $ab~ initio$ simulation package \cite{kresse1996_1,kresse1996_2} within
the general gradient approximation of Perdew-Burke-Ernzerhof type~\cite{perdew1996generalized}.
The cut-off energy for plane wave expansion is set to 500 eV.
In the self-consistent process, the $k$-point sampling grid of the BZ is 11 $\times$ 11 $\times$ 11.
The internal atomic positions are fully relaxed until the residual forces on each atom are less than
0.001 eV/\AA. 
In view of the small intrinsic SOC strength of Li, Mg and Si, SOC interaction is neglected in our calculations presented in the main text.

\subsection{The lattice and reciprocal vectors of SG 220}
\label{sup:B}
In a cubic body-centered structure of SG 220 with a lattice parameter $a$,
primitive lattice vectors ($\bt_1$, $\bt_2$, $\bt_3$) and primitive reciprocal vectors ($\bg_1$, $\bg_2$, $\bg_3$) are given in Cartesian coordinates as follows:
\begin{eqnarray}
&  \bt_1= (-a/2, a/2, a/2); ~~~~~~  \bg_1= (0, 2\pi/a, 2\pi/a);\notag \\ 
&  \bt_2= ( a/2,-a/2, a/2); ~~~~~~  \bg_2= (2\pi/a, 0, 2\pi/a);\\
&  \bt_3= ( a/2, a/2,-a/2); ~~~~~~  \bg_3= (2\pi/a, 2\pi/a, 0).\notag
\end{eqnarray}

\subsection{Representation matrices of the generators at H}
\label{sup:C}
Under the basis of the H4H5 irreps, the three unitary generators of the H point and 
the anti-unitary operator TRS
can be expressed~\cite{bradleybook} as
\beq
\begin{split}
& P:~\{IC^{-1}_{4x}|\frac{1}{2},0,0\}=iW\oplus -iW 
,~W\equiv
\left(
\begin{array}{ccc}
 0 & 1 & 0 \\
 -1 & 0 & 0 \\
 0 & 0 & -1 \\
\end{array}
\right);~ \\
& Q:~\{IC_{2,110}|\frac{1}{2},0,0\}=iX\oplus -iX 
,~X\equiv
\left(
\begin{array}{ccc}
 1 & 0 & 0 \\
 0 & 0 & 1 \\
 0 & 1 & 0 \\
\end{array}
\right);~ \\
& R:~\{C^{-1}_{3,1\bar1\bar1}|1,\frac{1}{2},\frac{1}{2}\}=A\oplus A
,~~~~~A\equiv
\left(
\begin{array}{ccc}
 0 & 0 & 1 \\
 1 & 0 & 0 \\
 0 & 1 & 0 \\
\end{array}
\right); \\
& {\cal T}:~TRS=E\otimes 
\left(
\begin{array}{ccc}
0&1 \\
1&0\\
\end{array}
\right){\cal K}
,~~~E\equiv
\left(
\begin{array}{ccc}
 1 & 0 & 0 \\
 0 & 1 & 0 \\
 0 & 0 & 1 \\
\end{array}
\right)
\end{split}
\eneq
where ${\cal K}$ denotes the complex conjugation. These operators (${\cal O}=P, Q, R, \cal T$) acting in momentum space are given by ${\cal O}[k_x,k_y,k_z]^T =g_{\cal O}[k_x,k_y,k_z]^T$ 
\beq
g_P=
\left(
\begin{array}{ccc}
-1 & 0 & 0 \\
 0 & 0 &-1 \\
 0 & 1 & 0 \\
\end{array}
\right),~
g_Q=
\left(
\begin{array}{ccc}
 0 &-1 & 0 \\
-1 & 0 & 0 \\
 0 & 0 & 1 \\
\end{array}
\right),~
g_R=
\left(
\begin{array}{ccc}
 0 &-1 & 0 \\
 0 & 0 & 1 \\
-1 & 0 & 0 \\
\end{array}
\right),~
g_{\cal T}=
\left(
\begin{array}{ccc}
-1 & 0 & 0 \\
 0 &-1 & 0 \\
 0 & 0 &-1 \\
\end{array}
\right)~
\eneq

\newpage
\subsection{Band dispersions of the low-energy effective $\bk\cdot \bp$ model}
\label{sup:D}

By fitting with the first-principles calculations, the parameters $\alpha$, $p$ and $q$ in the effective $\bk\cdot \bp$ Hamiltonian $\alpha\delta \bk^2+{\cal H}(\delta \bk)$ are obtained to be 12 eV$\cdot$\AA$^2$, 0.2 eV$\cdot$\AA~and 2.2 eV$\cdot$\AA, respectively. The dispersions in the [111] direction are shown in Fig. \ref{fig3}(c) in the main text, while the dispersions in the [100] and [110] directions are presented in Fig.~\ref{figs1}. 
One can find that the dispersions from our $\bk\cdot \bp$ model agree well with those from the first-principles calculations.

\begin{figure}[!h]
\includegraphics[width=3.5in]{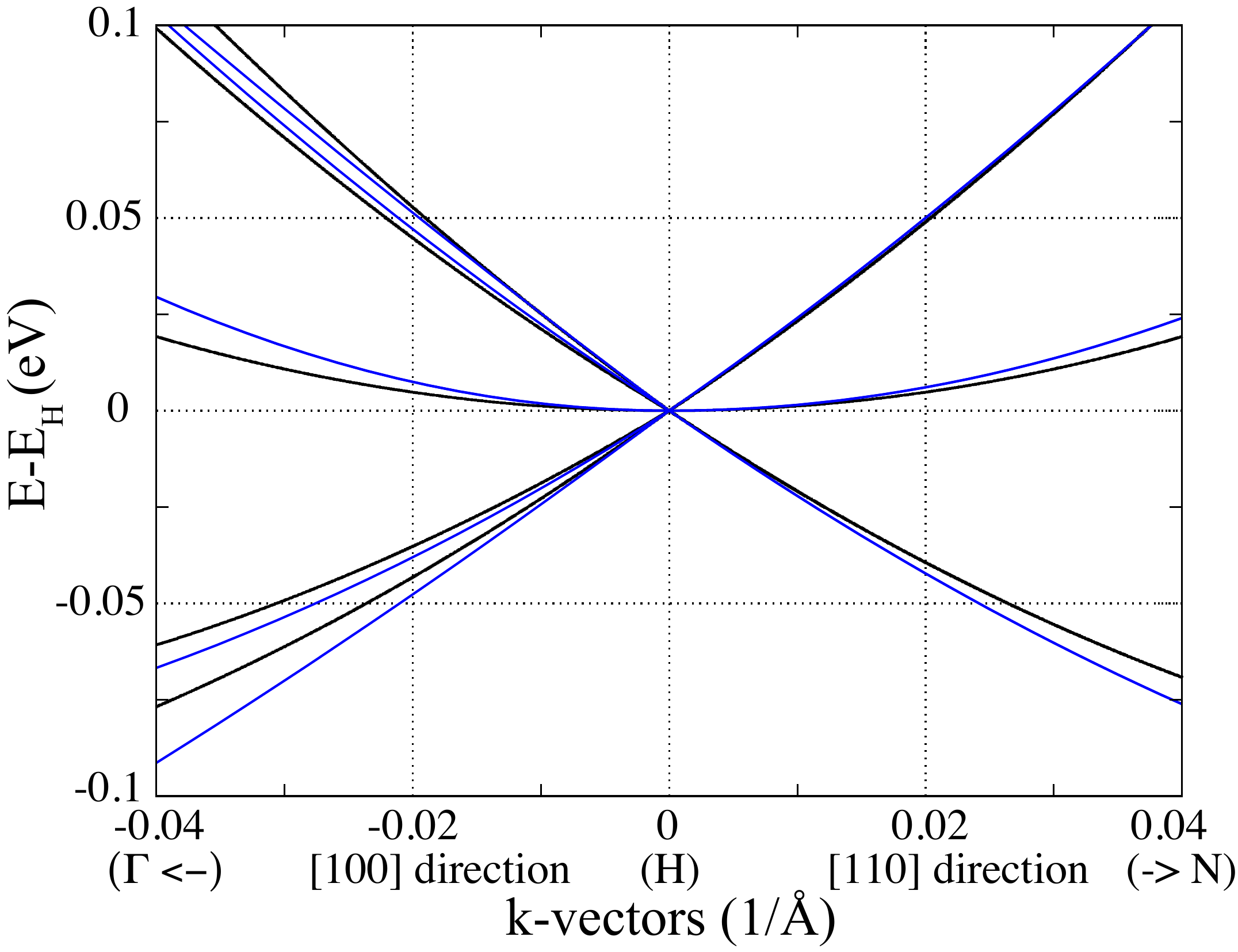}
     \caption{(color online). The band structures of the low-energy effective model along $\Gamma \gets \text{H}\to \text{N}$ with $\alpha=12~\text{eV}\cdot$\AA$^2$, $p=0.2~\text{eV}\cdot$\AA, $q=2.2~\text{eV}\cdot$\AA. The black and blue lines are obtained from the $\bk \cdot \bp$ model and first-principles calculations, respectively. The energy of the six-fold excitation at H is chosen as the reference energy (E$_\text{H}=0$ eV). 
}
\label{figs1}
\end{figure}

\subsection{The Wilson bands of 26 valence bands}
\label{sup:E}
Although the six-fold excitation is similar to the four-fold Dirac fermion in the sense that the total winding number of the two lowest bands is zero, the nontrivial winding number of each band is protected and robust. In the calculations of the Wilson loops, we also consider the additional 24 occupied bands in the energy range $-6$~eV to $-0.6$~eV dominated by $p$ orbitals of Si. The results are shown in Fig. \ref{figs2}. As discussed in the main text, the crossings of winding Wilson bands at $\varphi=\pi$ and $\theta=equ.$ are stable and protected by $\overline C_{2z}$ and ${\cal T}\wt M_{110}$, respectively.
In conclusion, the winding feature of $\pm 4\pi$ is robust against adding more trivial occupied bands in \lms.

\begin{figure}[!b]
\includegraphics[width=6.5in]{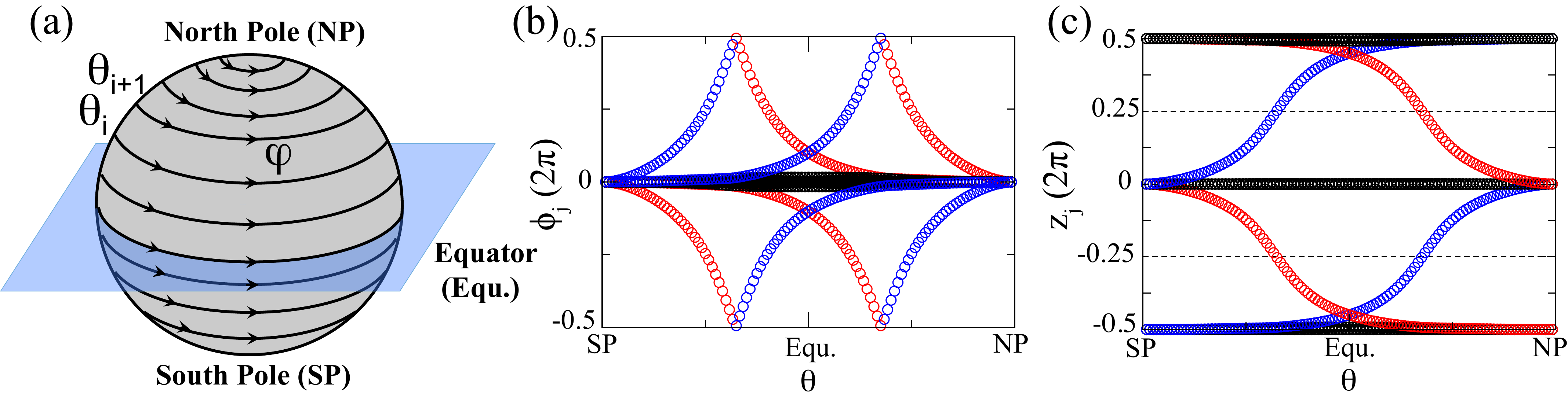}
     \caption{(color online). Wilson loop spectra for 26 occupied bands.
(a) A series of 1D Wilson loops on a closed sphere enclosing the six-fold excitation. (b) The Wannier charge centers for $\theta_j$ loops schematically shown in (a).
(c) The phases of the $\overline C_{2z}$ eigenvalues ($e^{-iz_j}$) for the Wilson bands in (b).
}
\label{figs2}
\end{figure}

\clearpage
\subsection{The TB Hamiltonian, surface states and strain effect}
\label{sup:F}

A six-band TB Hamiltonian with $s$-like orbitals at $12a$ Wyckoff sites is constructed. The $12a$ Wyckoff sites in SG 220 have six nonequivalent positions, \ie (3/8, 0, 1/4), (1/8, 0, 3/4), (1/4, 3/8, 0), (3/4, 1/8, 0), (0, 1/4, 3/8) and (0, 3/4, 1/8) [given in Cartesian coordinates in units of $a$]. An $s$-like orbital is considered on each position. 
Thus, the dimension of the TB Hamiltonian is six.
Considering the nearest- ($t_1$), next-nearest- ($t_2$) and next-next-nearest-neighbor ($t_3$) hoppings in the Slater-Koster method, the TB Hamiltonian is written in Eq. [\ref{tb}].
With the parameters given in Table \ref{table:tb}, a single six-fold excitation emerges
at the H point in the Hamiltonian, whose dispersions in all directions are
qualitatively consistent with the first-principles calculations. This Hamiltonian  can well reproduce the main feature of the six-fold excitation in \lms, as shown in Fig. \ref{fig:fig4}(a).
\begin{table}[!h]
\caption{The parameters (in units of eV) for the TB Hamiltonian.
}\label{table:tb}
\begin{tabular}{p{0.6cm} p{1cm}|p{0.6cm} p{1cm}|p{0.6cm} p{1cm}|p{0.6cm} p{1cm}}
\hline
\hline
 $\epsilon_0$:&0.2 &
$t_1$:&         0.2&
$t_2$:&        $-0.15$ &
$t_3$:&        $-0.02$ \\
\hline
\hline
\end{tabular}
\end{table}

Based on the TB Hamiltonian, we study the phase transitions through strain engineering, and 
calculate the surface states without and with strains. For the strain-free \lms, the six-fold excitation usually leads to 
`Fermi-arc' states on the surfaces although they are not topologically protected. 
Like the `surface arcs' in the Dirac semimetals, the starting point and ending point coincide for the `Fermi-arc' states in \lms~on the (001)-surface [Fig.~\ref{figs3}(a)] and (110) surface [Fig.~\ref{figs4}(a)].
Next, we consider the phase transitions in \lms~through strain engineering 
since the state with the six-fold excitation is located at the phase boundary. 
It can be driven into various topological phases, such as Weyl semimetal and nodal-line semimetal, {\it etc}. A phase transition from the six-fold excitation to four Weyl points on the $k_z = 0$ plane can be triggered by the [001] strain. 
In spite of the short distances between the four split Weyl points, 
they give rise to two long `Fermi-arc' states terminated on their projections for both (001) [Figs.~\ref{figs3}(b,c)] and (110) surfaces [Figs.~\ref{figs4}(b,c)].
The surface `Fermi arcs' states are topologically protected by the chiral charges of the Weyl points, 
and confirm the topological phase transition through the [001] strain.
\begin{figure}[!h]
\includegraphics[width=4.2in]{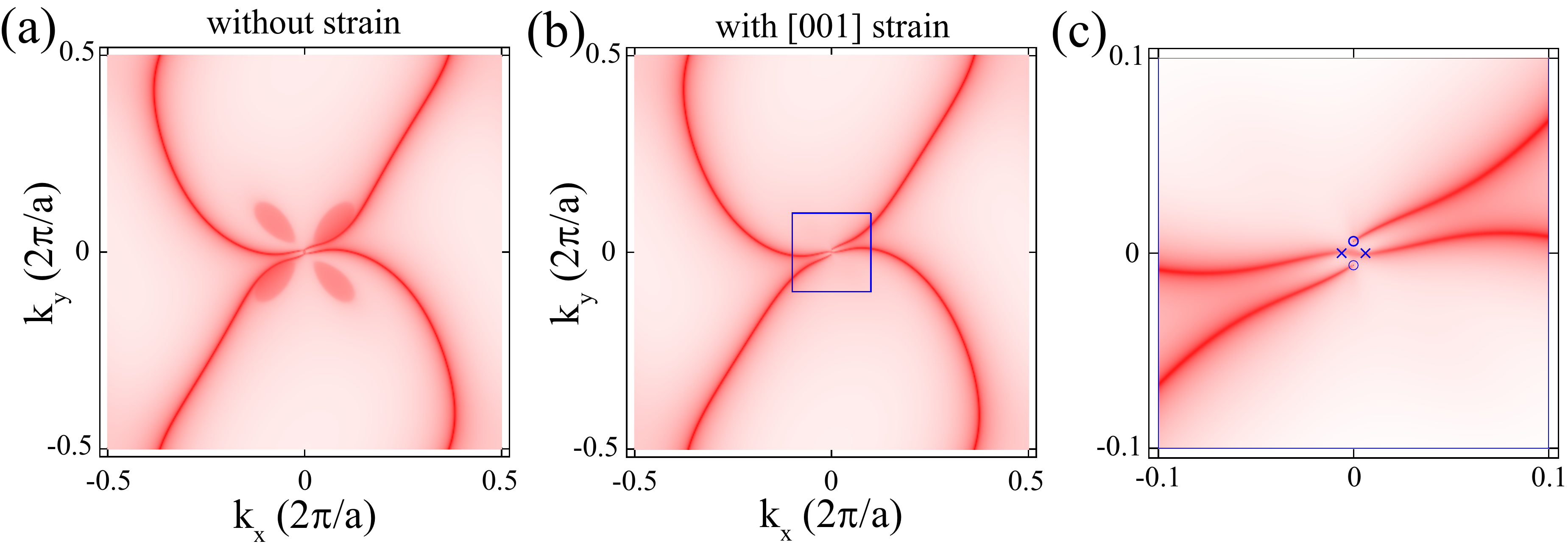}
\caption{(color online). 
The (001)-surface constant energy contours ($E=-0.007$ eV) for strain-free (a) and  [001]-strained (b, c) Hamiltonians. The compressive strain is applied along the [001] direction. Panel (c) is the zoom-in plot of the boxed area in panel (b). The ``x" and ``o" symbols stand for the Weyl points with opposite chirality.
}
\label{figs3}
\end{figure}

\begin{figure}[!h]
\includegraphics[width=5.2in]{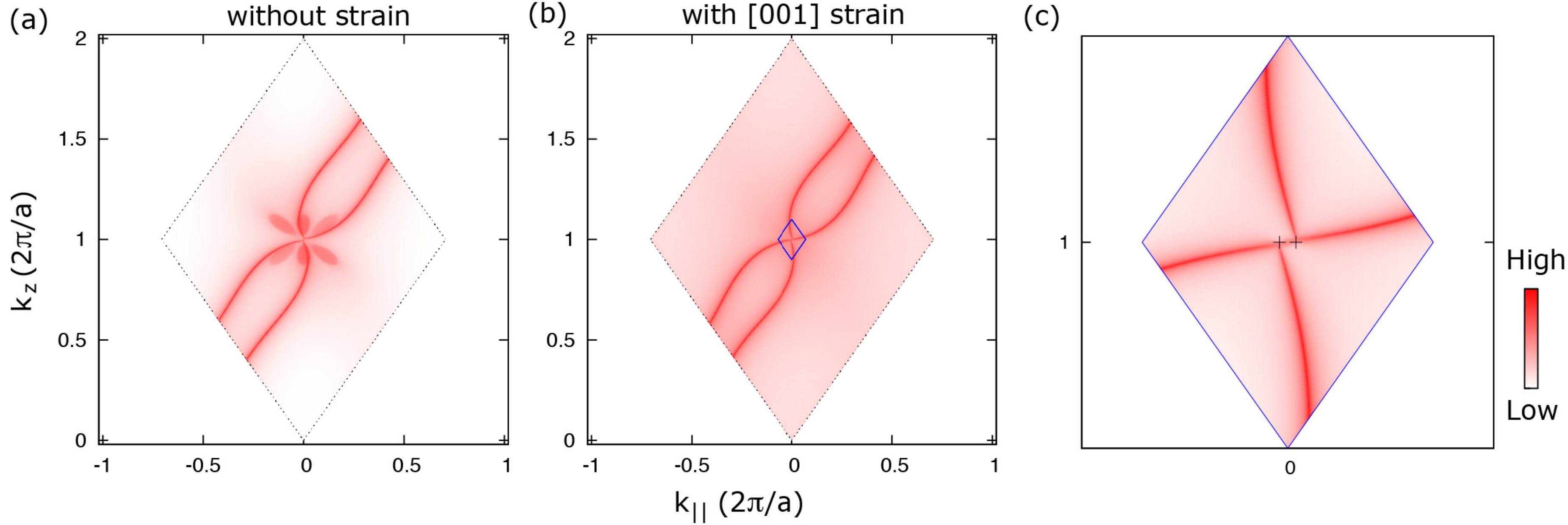}
\caption{(color online). 
(110)-surface constant energy contours ($E=-0.007$ eV) for strain-free (a) and  [001]-strained (b, c) Hamiltonians. The compress strain is along the [001] direction. Panel (c) is the zoomin plot of the boxed area in panel (b). The ``+" symbols stand for the projections of the Weyl points.
}
\label{figs4}
\end{figure}

\newpage

In addition, the six-fold excitation evolves into two sets of nodal lines upon applying the [111] strain, as shown in Fig.~\ref{figs5}(a), which can be related by TRS.
Each set of the nodal lines consists of four segments [Figs.~\ref{figs5}(b,c)], which are protected by the three-fold rotation about the [111] direction, $\wt M_{1-10}$, $\wt M_{-101}$ and $\wt M_{01-1}$, respectively.  

\begin{figure}[!h]
\includegraphics[width=5.8in]{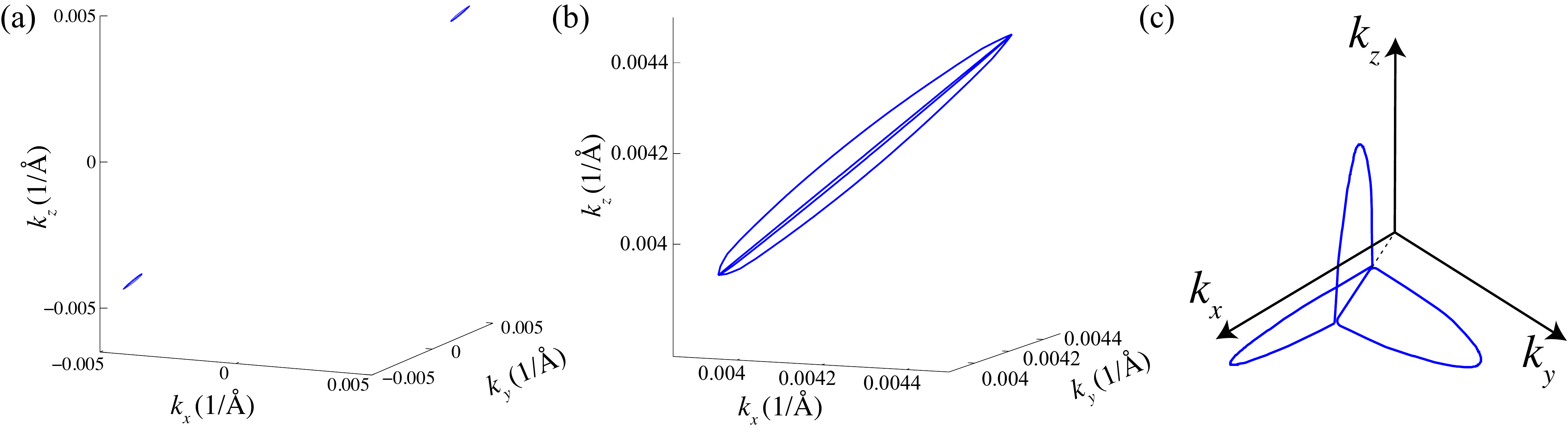}
\caption{(color online). 
The nodal lines in [111]-strained Hamiltonian. (a) Two sets nodal lines derived from a single six-fold excitation. (b) and (c) Two different views 
of one set of the nodal lines. One set of nodal lines consists of one line segment in [111] direction and three $C_{3,111}$-related arc segments.
}
\label{figs5}
\end{figure}

\subsection{Hinge states}
\label{sup:G}
In contrast to the topologically trivial surface `Fermi-arc' states, all the energetically-gapped $k_z$ slices belong to a 2D HOTI phase, which results in topological arc states on the [001] hinges of \lms. The topologically nontrivial phase is closely related to a filling anomaly: a mismatch between the number of electrons required to simultaneously satisfy charge neutrality and the crystal symmetry (${\cal T}\wt S_{4z}$). 
By computing 1D Wilson loops along the $\bb_1^*$ ($\equiv\bg_3$) and $\bb_2^*$ ($\equiv \bg_1-\bg_2$) directions, our results show that the two charge centers are quantized to [0.25$\bb_1$, 0.25$\bb_2$] and [0.75$\bb_1$, 0.75$\bb_2$], which correspond to $\{0, 0.25a\}$ and $\{0, 0.75a\}$, respectively, in Cartesian coordinates in the projected $xy$-plane [Fig.~\ref{figs6}(a)]. The 1D Wilson loops are along the $\bb_1^*$ direction and the results are shown in Fig.~\ref{figs6}(b). The nested Wilson loops are along the $\bb_2^*$ direction and the results are shown in Fig.~\ref{figs6}(c) for two individual Wilson bands (well-separated) in Fig.~\ref{figs6}(b). The quantizations are protected by the ${\cal T}\wt S_{4z}$ symmetry (with $\wt S_{4z}\equiv\{IC^{-1}_{4z}|00\frac{1}{2}\}$) for each energetically-gapped 2D $k_z$-slice. Because the two obtained Wannier charge centers are two distinct ${\cal T}\wt S_{4z}$-invariant positions, it is impossible to have any choice of Wannier center assignment that preserves charge neutrality and $\wt S_{4z}$ symmetry simultaneously. 
\begin{figure}[!h]
\includegraphics[width=5.8in]{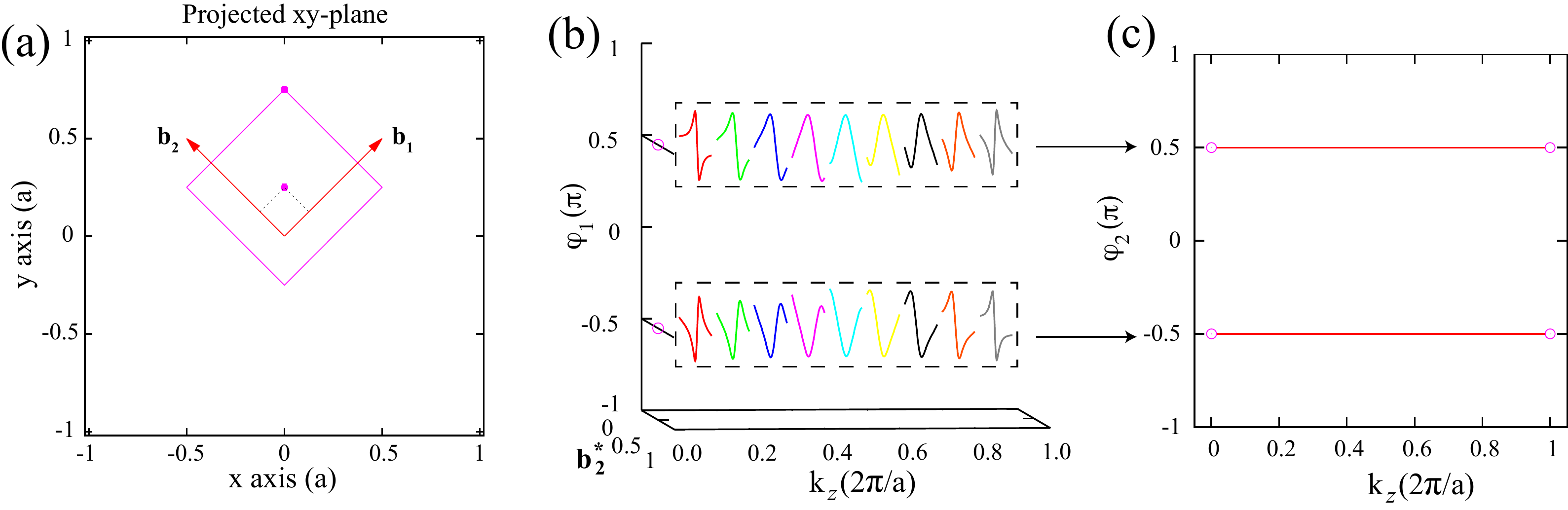}
\caption{(color online). 
The Wannier charge centers of the two lowest bands of the sixfold excitation in \lms. (a) The definition of the origin, $\bb_1$ and $\bb_2$ in the projected $xy$-plane in real space. The orange-colored box is $\wt S_{4z}$-invariant. (b) The Wannier charge centers of the $\bb_1^*$ directed Wilson loops as a function of $\bb_2^*$ and $k_z$. Due to the gapless point of the sixfold excitation, the two lowest bands on the loop (\ie $\bb_2^* = \pi$, $k_z = 0$) is ill-defined, which is denoted by circles. (c) The Wannier charge centers of the nested Wilson loops along the $\bb_2^*$ direction as a function of $k_z$. Due to the gapless point of the sixfold excitation, the two lowest bands on the $k_z = 0$ plane is ill-defined, which is denoted by circles. The $\bb_1^*$ is defined as the $\bf{g}_3$ direction (i.e., [110] direction), while $\bb_2^*$ is defined as the $\bf{g}_1-\bf{g}_2$ direction (i.e., [$\bar{1}$10] direction).  The $k_z$ is defined as the $\bf{g}_1+\bf{g}_2-\bf{g}_3$ direction, which is the $z$-axis.
}
\label{figs6}
\end{figure}

Since the existence of hinge arc states is one of the hallmarks of the six-fold excitation, we build a 1D $z$-directed rod with a periodic boundary condition in $z$ direction, and open boundary conditions along both $\bb_1$ and $\bb_2$ directions. The size of the supercell is 20 primitive cells in both $\bb_1$ and $\bb_2$ directions, \ie a $20\times20$ supercell. For this purpose, we redefine the origin of the primitive cell as $\{0,-0.25\}$ in Fig.~\ref{figs6}(a) and the three primitive lattice vectors as: $\bt_3$, $\bt_1$ and $\bt_1+\bt_2$.  The band structure of the $\wt S_{4z}$-symmetric rod is shown in Fig. \ref{figs7}(a). It is clear that there are four nontrivial hinge states (blue-colored) in the band gap of bulk states (black-colored), whose starting point and ending point are the six-fold excitation.  
In order to clearly show the characteristic of these four hinge states, we calculate their real-space distributions on the rod, as shown in Figs. \ref{figs7}(b-g). When $k_z=\frac{\pi}{a}$, the hinge states are well-isolated from the bulk states in energy space and 
weakly coupled to the bulk states, leading to that they are localized at the four hinges of the rod, as shown in Fig. \ref{figs7}(g).
When smaller $k_z$ is chosen, the hinge and bulk states are closer in energy space and the couplings between them become stronger.
At last, the hinge states with $k_z$ close to 0 (\ie $\wt \Gamma$) are very extended in the real space. Therefore, the presence of the nontrivial hinge states confirms the nontrivial bulk topology in \lms.

\begin{figure}[h]
\includegraphics[width=6.5in]{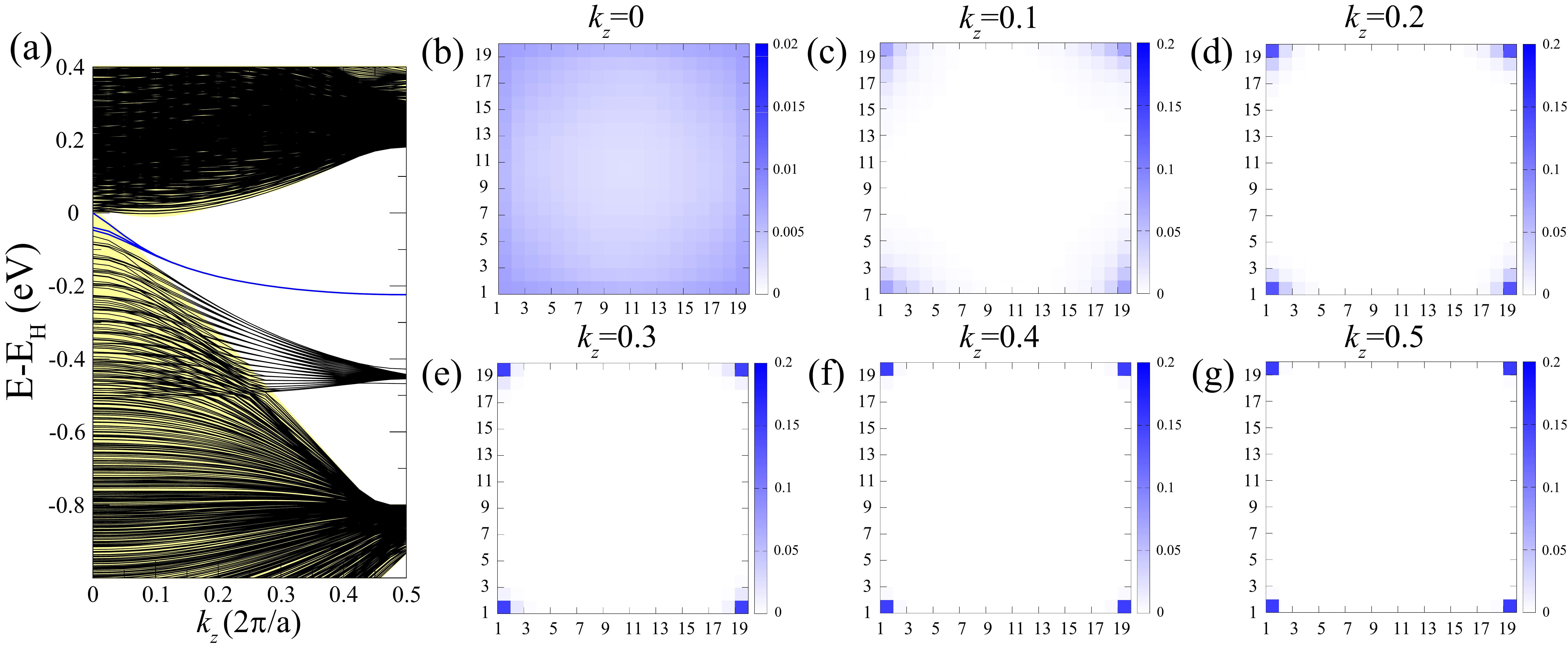}
\caption{(color online). 
(a) The [001] hinge dispersions of \lms~in a larger energy range compared with the dispersions in Fig. \ref{fig:fig4}(c). 
 The real-space distributions of the hinge states for $k_z=0$ (b), $k_z=0.1$ (c), $k_z=0.2$ (d), $k_z=0.3$ (e), $k_z=0.4$ (f), and $k_z=0.5$ (g) in units of $\frac{2\pi}{a}$, on the structure of a 1D $20\times20$ supercell. The hinge states and bulk states are colored in blue and black, respectively.
}
\label{figs7}
\end{figure}

\subsection{Band structure of \lms~with SOC and $A_{12}A'_{3}B_{4}$ without SOC}
\label{sup:H}

We also calculate the band structure of \lms~with SOC, as shown in Fig.~\ref{figs8}.
The obtained SOC band gap is so tiny ($\sim$ 0.7 meV from the inset of Fig.~\ref{figs8}), indicating the negligible strength of SOC.
Once the spin degrees of freedom is considered, the six-fold excitation becomes a nearly twelve-fold excitation in \lms. 
Therefore, it is promising to find the twelve-fold excitation in \lms.

\begin{figure}[h]
\includegraphics[width=3.5in]{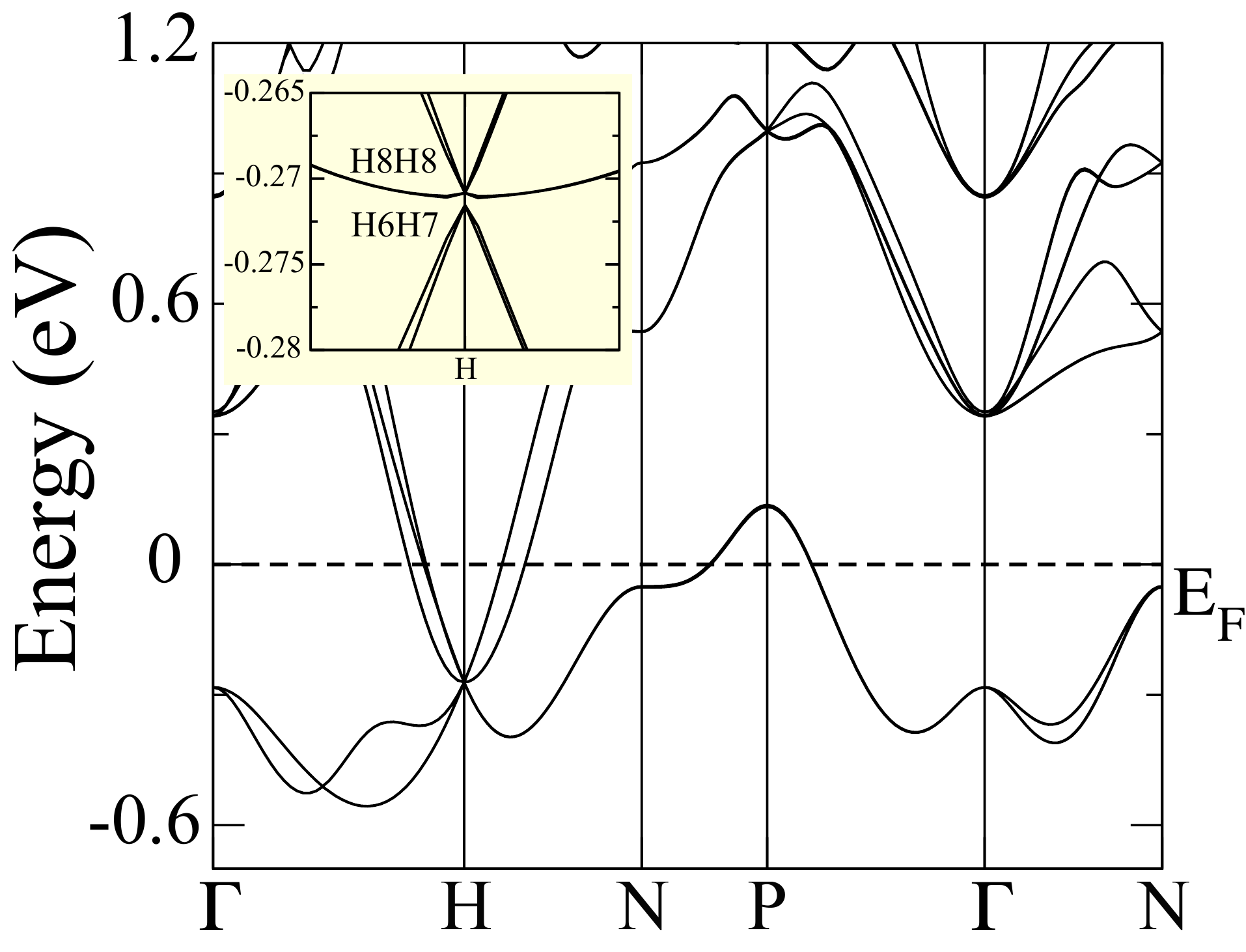}
      \caption{(color online). The band structure of \lms~with SOC.
      The inset shows the zoom-in band structure around the H point with a tiny band gap.
The twelve-fold excitation splits into a lower four-fold irrep of H6H7 and a higher eight-fold irrep of H8H8.
The program ``CheckTopologicalMat" tells that it is a symmetry-enforced semimetal with a tiny SOC gap at H.
It's a symmetry-enforced semimetal.
}
\label{figs8}
\end{figure}

\begin{figure}[t]
\includegraphics[width=6.2in]{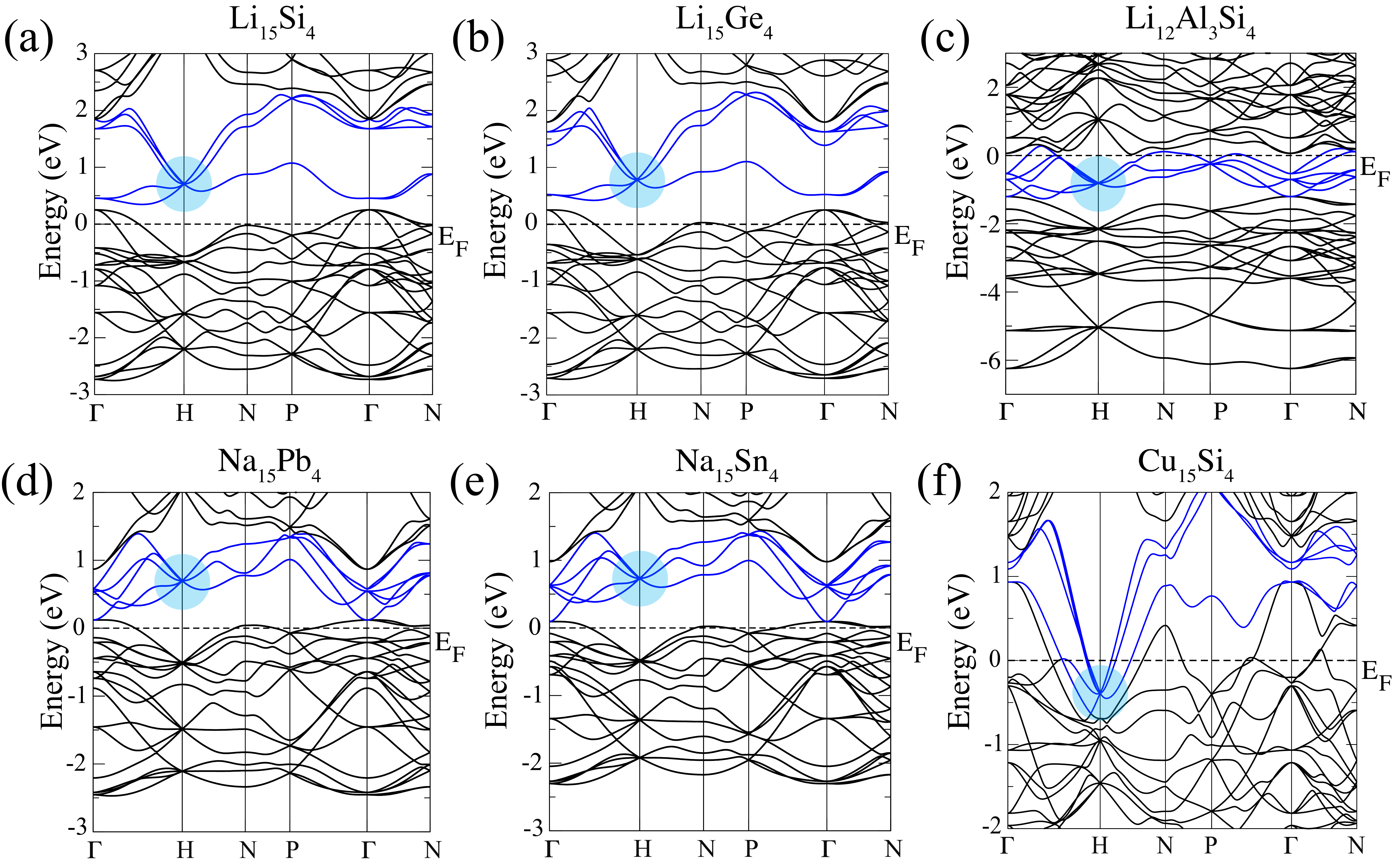}
      \caption{(color online). The band structures of the synthesized crystals Li$_{15}$Si$_4$ (a), Li$_{15}$Ge$_4$ (b), Li$_{12}$Al$_3$Si$_4$ (c), Na$_{15}$Pb$_4$ (d), Na$_{15}$Sn$_4$ (e) and Cu$_{15}$Si$_4$ (f) without SOC. The six-fold excitation is highlighted in a shadowed circle.
}
\label{figs9}
\end{figure}

In addition to \lms, its derivatives $A_{12}A'_{3}B_{4}$ ($A=$Li, Na, Cu; $A'$=Li, Na, Mg, Al, Cu;  $B=$Si, Ge, Sn, Pb) are a large family of materials, which have been explored too. Although some of them haven't been synthesized, the electrides Li$_{15}$Si$_4$~\cite{zeilinger2013stabilizing}, Li$_{15}$Ge$_4$~\cite{johnson1965crystal}, Li$_{12}$Al$_3$Si$_4$~\cite{pavlyuk1992crystal}, Na$_{15}$Pb$_4$~\cite{lamprecht1968solubility},  Na$_{15}$Sn$_4$~\cite{baggetto2013characterization}, and Cu$_{15}$Si$_4$~\cite{weitzer1990phase} have been obtained.  
The band structures of these materials without SOC are shown in Fig. \ref{figs9}, which clearly show the existence of 
six-fold excitations around the Fermi level.

\newpage
\end{widetext}
\end{document}